\pdfoutput=1
\documentclass[article, shortnames, nojss]{jss}

\usepackage{lmodern}

\newcommand{\class}[1]{`\code{#1}'}
\newcommand{\fct}[1]{\code{#1()}}

\usepackage[utf8]{inputenc}
\showboxdepth=\maxdimen
\showboxbreadth=\maxdimen
\usepackage{float}
\usepackage{amsmath}
\usepackage{amssymb}

\newcommand{\Nystrom}{Nystr\"{o}m}
\newcommand{\Matern}{Mat\'{e}rn}
\newcommand{\R}{\mathbb{R}}
\newcommand{\D}{\mbox{${\cal D}$}}

\newcommand{\uf}[1]{\noindent {\bf #1}}
\newcommand{\ve}[1]{\mbox{\boldmath ${#1}$}}
\newcommand{\bm}[1]{\ve{#1}}
\newcommand{\vesub}[2]{\ve{#1}_{#2}}
\newcommand{\vesup}[2]{\ve{#1}^{#2}}
\newcommand{\vess}[3]{\ve{#1}_{#2}^{#3}}

\newcommand{\hvesub}[2]{\hat{\ve{#1}}_{#2}}
\newcommand{\hvesup}[2]{\hat{\ve{#1}}^{#2}}

\newcommand{\tr}{\mbox{tr}}

\setkeys{Gin}{width=0.5\textwidth}

\author{Evandro Konzen \\ Newcastle University
   \And Yafeng Cheng \\ Newcastle University
   \And Jian Qing Shi \\ Southern University of \\ Science and Technology
   }
\Plainauthor{Evandro Konzen, Yafeng Cheng, Jian Qing Shi}

\title{Gaussian Process for Functional Data Analysis: The \pkg{GPFDA} Package for \proglang{R}}
\Plaintitle{Gaussian Process for Functional Data Analysis: The GPFDA Package for R}
\Shorttitle{\pkg{GPFDA}: Gaussian Process for Functional Data Analysis}

\Abstract{
We present and describe the \pkg{GPFDA} package for \proglang{R}. The package provides flexible functionalities for dealing with Gaussian process regression (GPR) models for functional data. Multivariate functional data, functional data with multidimensional inputs, and nonseparable and/or nonstationary covariance structures can be modeled. In addition, the package fits functional regression models where the mean function depends on scalar and/or functional covariates and the covariance structure is modeled by a GPR model. In this paper, we present the versatility of \pkg{GPFDA} with respect to mean function and covariance function specifications and illustrate the implementation of estimation and prediction of some models through reproducible numerical examples.
}

\Plainkeywords{covariance function, Gaussian process regression, functional regression, nonseparable, nonstationary, multivariate, R}

\Address{
  Evandro Konzen, Yafeng Cheng\\
  School of Mathematics, Statistics and Physics\\
  Newcastle University\\
  Newcastle upon Tyne, United Kingdom\\
  
  Jian Qing Shi\\
  Department of Statistics and Data Science\\
  Southern University of Science and Technology\\
  Shenzhen, China\\
  E-mail: \email{shijq@sustech.edu.cn}
}

\begin{document}




\section[Introduction]{Introduction} \label{sec:intro}

Functional data analysis (FDA) has been an active research area in the modeling of a variety of types of data, such as curves, images, and spatial and spatiotemporal data. In this area, data are seen as discretely observed realizations of a continuous stochastic process. Furthermore, relationships between functional variables and relationships between functional variables and other quantities can be analyzed through functional regression (FR) models. A general overview of recent advances in FDA can be found in \cite{wang2016fda}. More detailed discussion focused on FR models is given by \cite{greven2017general}. Some overview on functional spatial data analysis is provided in \cite{delicado2010spatfd} and \cite{giraldo2018spatialbigdata}.

Modeling functional data with multidimensional inputs (or covariates) is often a challenging task. While parametric approaches are usually inflexible and restricted to special cases, nonparametric approaches commonly face the well-known \textit{curse of dimensionality}. These difficulties can be aggravated when there is a multivariate response. To tackle these difficulties, functional data and FR models can be addressed by Gaussian process regression (GPR) models \citep{shi2007batch, shi2011gaussian}, where functional data are seen as realizations from a Gaussian process (GP) with a covariance kernel from a known parametric family.

This paper describes the \pkg{GPFDA} package \citep{GPFDA} for \cite{Rcoreteam} for GPR models for functional data. 
The current version of \pkg{GPFDA} provides functionalities to work with the following models: univariate GP regression with either (i) stationary and separable covariance structure (GPR) or (ii) nonstationary and/or nonseparable covariance structure (NSGPR); multivariate GPR (MGPR); and GP functional regression (GPFR). In the next paragraphs, we give an overview of other \proglang{R} packages that can be used for working with similar models.

Many available \proglang{R} packages for GPR models are restricted to the case of stationary covariance functions, e.g., \pkg{DiceKriging} and \pkg{DiceOptim} \citep{DiceKrigingDiceOptim}, \pkg{spatial} \citep{spatial}, \pkg{gstat} \citep{gstat}, and \pkg{geoR}  \citep{geoR}. \pkg{GPFDA} package is not intended to overcome all these packages in all aspects under the stationary context, but rather to provide flexibility in modeling. Nevertheless, even in the stationary case \pkg{GPFDA} does offer the use of different mean function specifications and multiple independent realizations to learn the covariance structure. 

Computational time is not the primary aim of \pkg{GPFDA}, as it is for the \proglang{R} packages \pkg{bigGP} \citep{bigGP} and \pkg{laGP} \citep{laGP}. Nevertheless, \pkg{GPFDA} does use efficient \proglang{C++} code, offers the use of analytical gradient for several covariance kernels, and provides options of approximation methods if computational time is a concern. We also include Subset of Data as learning approximation method and Subset of Regressors (similar to \Nystrom{} method) for prediction (a discussion about these methods can be
seen in \cite{rasmussen2006gaussian}).

Some \proglang{R} packages allow for nonstationary covariance functions, but are restricted to specific families, e.g., \pkg{RandomFields} \citep{RandomFields}. Other examples introduce nonstationarity by partitioning the space into regions and fitting stationary GPR model separately within each region (see \pkg{tgp} \citep{tgp}), or by using local maximum likelihood (see \pkg{convoSPAT} \citep{convoSPAT}). Alternatively, \pkg{GPFDA} employs a flexible nonstationary and/or nonseparable model for the covariance function using B-spline representation for the time and/or spatially varying parameters following \cite{konzen2020modelling}. This modeling approach can easily be applied to input dimensions larger than two and does not require the difficult choices of partition of the input space.

GPR for multivariate responses can be dealt by the \proglang{R} package \pkg{mlegp} \citep{mlegp}, which fits independent GPs to each dimension. To consider cross-covariance structure, \pkg{GPFDA} also deals with multivariate responses through the MGPR model that is based on convolution processes. This model is proposed by \cite{boyle2004dependent} and further discussed by \cite{shi2011gaussian}, and is an alternative approach 
to coregionalization models used by the \proglang{R} package \pkg{spBayes} \citep{spBayes}. 
In MGPR, the covariance structure of the multivariate response is defined by 
latent processes which describe the individual behavior of each output and latent 
processes which model the pairwise dependence between outputs. 

Finally, \pkg{GPFDA} can be used for the GPFR model, where the mean function may depend on scalar and/or functional covariates and the covariance structure is defined by a GP which may itself depend on functional covariates. The mean function is estimated by a functional regression (FR) model and the covariance structure is estimated by a GPR model. As far as we know, up to the present date, there is no \proglang{R} package which uses GPs within a functional regression model.

The implementation of the FR part in GPFR models in \pkg{GPFDA} 
follows \citet{ramsay2005functional} and can include a mix of scalar and functional covariates.
Similar implementations are done by \pkg{fda} \citep{fda} and \pkg{fda.usc} \citep{fda.usc} packages. 
Other packages for function-on-scalar and function-on-function regression are 
available on CRAN. They include methods for sparsely or densely sampled random 
trajectories -- \pkg{fdapace} \citep{fdapace}; spline-based methods for penalizing 
roughness -- \pkg{refund} \citep{refund}; and additive regression models and 
variable selection -- \pkg{FDboost} \citep{FDboost}.

The remainder of the paper is organized as follows. Section~\ref{sec:meth} gives an overview of the methodology used in the package. Section~\ref{sec:package} describes the main functionalities of the package for each model. Some examples are illustrated in Section~\ref{sec:examples} and future extensions of the package are discussed in Section~\ref{sec:concl}. Computational details are mentioned in the last section.

\section[Methodology]{Methodology} \label{sec:meth}

\subsection[Univariate Gaussian process regression (GPR)]{Univariate Gaussian process regression (GPR)} \label{subsec:GP}

 Let $x$ be a functional variable and $\ve t$ be a $Q$-dimensional covariate. A nonparametric regression model is expressed as
\[
x = f(\ve t) + \epsilon, \ \ \epsilon \ \sim  \ N(0,\sigma_\epsilon^2),
\]
where $f(\cdot)$ is unknown. However, most of the nonparametric methods suffer from
the {\it curse of dimensionality} when they are applied to the problem with multi-dimensional
covariates (i.e., $Q$ is large). A variety of alternative approaches has been developed to
overcome this problem. Examples include the additive model \citep{breiman1985additive}, the projection pursuit regression \citep{friedman1981pursuit}, the sliced inverse regression \citep{li2009sliced}, the neural network model \citep{cheng1994neural}, the varying-coefficient model \citep{hastie1993varying, fan1999varying} and the GPR model \citep{ohagan1978curvefitting}.

The GPR model is a nonparametric model and has some nice features; see details in  
 \cite{shi2011gaussian}. Suppose we have a data set
\[
\D = \left \{ \left(
\begin{array}{c}
x_1\\
{\ve t_1} \\
\end{array}
\right) ~
\left(
\begin{array}{c}
x_2\\
{\ve t_2} \\
\end{array}
\right) \ldots
\left(
\begin{array}{c}
x_n\\
{\ve t_n} \\
\end{array}
\right) \right \}.
\]
The discrete form of a GPR model is defined as follows. 
\begin{eqnarray}
x_i &= & f(\vesub{t}{i}) + \epsilon_i,~i=1,\ldots,n, \label{eq:2.basicm} \\
  \epsilon_i & \sim & i.i.d. \ N(0,\sigma_\epsilon^2),  \nonumber \\
  f(\cdot ) & \sim & GP(\mu(\cdot),k(\cdot,\cdot))~\mbox{and}~{\COV}(f(\vesub{t}{i}),f(\vesub{t}{j}))=k(\vesub{t}{i},\vesub{t}{j}), \nonumber
\end{eqnarray}
where $GP(\mu(\cdot),k(\cdot,\cdot))$ is a GP prior with mean function $\mu(\cdot)$ and covariance function $k(\cdot,\cdot)$. GP here can be treated as a prior of the unknown function $f(\cdot)$ from a Bayesian viewpoint.

Stationary covariance functions used in \pkg{GPFDA} package can be seen in Table~\ref{tab:statcovfuns}. The \Matern, powered exponential and rational quadratic models are functions of the (squared) distance given by
\begin{equation}
d_{(\gamma)} = \sum^Q_{q=1} w_q |\ve t_{q}-\ve t_{q}'|^\gamma, \qquad \omega_q \geq 0,  \quad   0 < \gamma \leq 2.
\end{equation}
The powered exponential is also known as \textit{exponential} when $\gamma=1$ and \textit{squared exponential} when $\gamma=2$.

\renewcommand\arraystretch{1.2}
\begin{table}[t!]
\centering
\begin{tabular}{lp{11cm}}
\hline
Model & Covariance function $k(\ve t, \ve t')$ \\ \hline
Linear &	$a_0 + \sum^Q_{q=1}a_q \ve t_{q} \ve t_{q}'$ \\
\Matern &	$ k(\ve t, \ve t') = v_0 \frac{1}{\Gamma(\nu)2^{\nu-1}} \Big( \sqrt{2 \nu d_{(2)} } \Big)^{\nu} {\cal{K}}_{\nu} \Big( \sqrt{2 \nu d_{(2)}} \Big)$, where ${\cal{K}}_{\nu}$ is the modified Bessel function of order $\nu$. \\
Powered exponential &	$v  \exp (-d_{(\gamma)} )$, \\ 
Rational quadratic & $v (1 + d_{(2)})^{-\alpha}, \ \alpha \geq 0	$ \\ \hline
\end{tabular}
\caption{\label{tab:statcovfuns} Stationary covariance functions.}
\end{table}
\renewcommand\arraystretch{1}

The \Matern{} class is very general and can accommodate several particular cases. For example, as $\nu \rightarrow \infty$, the \Matern{} covariance function converges to a squared exponential one. In machine learning, we often encounter applications with using $\nu=3/2$ and $\nu=5/2$. This is because if $\nu=p + 1/2$, where $p$ is a non-negative integer, the resulting covariance function is a product of a polynomial of order $p$ and an exponential \citep{rasmussen2006gaussian}. If $\nu=1/2$, we obtain an equivalent expression to the exponential covariance function.

Note that we can use a combination of these covariance functions by taking the sum of them.

\subsubsection{Fitted values and predictions} \label{gpr:fitted}

We now temporarily assume that the noise variance $\sigma_\epsilon^2$ is known, and the covariance function $k(\cdot,\cdot)$ is predetermined with fixed hyper-parameters known in advance. We use $\ve \theta$ to denote $\sigma_\epsilon^2$ and the hyper-parameters. They can be estimated by using, for example, the empirical Bayesian approach which will be discussed below. It is also common to assume a zero mean function, i.e., $\mu(\cdot) = {0}$.

Let $\ve f =(f(\ve t_1), \ldots, f(\ve t_n))^\top$. When the value of the hyper-parameters $\ve \theta$ is given, the posterior distribution, $p(\bm{f}|\D,\sigma_\epsilon^2)$, is a multivariate normal distribution with
 \begin{eqnarray*}
   \E(\bm{f}|\D,\sigma_\epsilon^2)& =
    & \bm{K}  (\bm{K} + \sigma_\epsilon^2 \bm{I})^{-1} \ve{x},  \\
 \VAR(\bm{f}|\D,\sigma_\epsilon^2)& = & \sigma_\epsilon^2 \bm{K}  (\bm{K} + \sigma_\epsilon^2 \bm{I})^{-1},
\end{eqnarray*}
where the covariance matrix $\bm{K}$ is calculated by using the kernel covariance function. Its $(i,j)$th element is calculated by
\begin{equation}
\bm{K}(i,j)=\COV(f_i, f_j)=k(\vesub{t}{i}, \vesub{t}{j}; \ve{\theta}).  \label{eq:3.kij}
\end{equation}
Note that the mean vector of this GP prior is assumed to be zero.

It is straightforward to predict an output for new data points, i.e., the points other than $\{\vesub{t}{1}, \ldots, \vesub{t}{n}\}$. We also call them as \emph{test data} and call $\D$ as \emph{training data}. Let $\vesup{t}{*}$ be a new input and let
$f(\vesup{t}{*})$ be the related nonlinear function. The vector $(f(\vesub{t}{1}),\ldots,f(\vesub{t}{n}),f(\vesup{t}{*}))$ constitutes a $(n+1)$-variate normal vector. Consequently, the posterior distribution of $f(\vesup{t}{*})$ given the
training data $\D$ is also a Gaussian distribution, with
mean and variance given by
\begin{eqnarray}
\E(f(\vesup{t}{*})|{\D}) & = &  \bm{\psi}^\top(\vesup{t}{*})
\bm{\Psi}^{-1} \ve{y}, \label{eq:2.pr10} \\
\VAR(f(\vesup{t}{*})|\D) & = & k(\vesup{t}{*},\vesup{t}{*})
- \bm{\psi}^\top(\vesup{t}{*}) \bm{\Psi}^{-1}
\ve{\psi} (\vesup{t}{*}) \label{eq:2.pr20},
\end{eqnarray}
where  $\ve{\psi}(\vesup{t}{*})=(k(\vesup{t}{*}, \vesub{t}{1}),
\cdots, k(\vesup{t}{*}, \vesub{t}{n}))^\top$ is the covariance between \sloppy  ${f(\vesup{t}{*})}$ and ${\ve f=(f(\vesub{t}{1}),\ldots,f(\vesub{t}{n}))}$, and $\ve{\Psi}$ is the
covariance matrix of $(x_1, \cdots, x_n)$ given by
\begin{equation} \ve{\Psi}=\ve K + \sigma_\epsilon^2 \ve I.
\label{Psi}
\end{equation}

If $x^*$ is the related output or response to $\vesup{t}{*}$, then its predictive distribution is also Gaussian, with the mean given by (\ref{eq:2.pr10}) and the variance
\begin{equation}
\hat{\sigma}^{*2}=\VAR(f(\vesup{t}{*})|{\cal D}) + \sigma_\epsilon^2.
\label{eq:2.pr30}
\end{equation}

If we use the posterior mean $\E(f(\vesup{t}{*})|{\D})$ in \eqref{eq:2.pr10} as the prediction of $f(\ve t^*)$, it satisfies posterior consistency, i.e., it is a consistent estimator of the true function $f_0(\cdot)$ \citep{shi2011gaussian}.   

\subsubsection{Empirical Bayes estimates}

In Bayesian inference, we usually select the values of hyper-parameters based on our prior knowledge. We however should be cautious about doing so for the GPR model since the dimension of $\ve \theta$ is usually quite large and we do not usually know the meaning or physical interpretation of   $\ve \theta$.  An alternative way is to estimate $\ve \theta$ using the observed data. This is so called empirical Bayes estimates \citep{carlin2008bayesian, shi2011gaussian}. 

  Using empirical Bayesian approach, we estimate $\ve{\theta}$ from the marginal distribution of ${\ve{x}=(x_1, \ldots, x_n)^\top}$:
\begin{equation}
p(\ve{x}|\ve{\theta})=\int p(\ve{x}|\ve{f}) p(\ve{f}|\ve{\theta}) d \ve{f}, \label{eq:3.marg}
\end{equation}
where $p(\ve{x}|\ve{f}) = \prod_{i=1}^n g(f_i)$ and $\ve{f} \sim N(\ve{0}, \bm{K})$. The $(i,j)$th element of the covariance matrix $\bm{K}$ is calculated by   \eqref{eq:3.kij}.
Consequently, for the continuous response with normal distribution as given in \eqref{eq:2.basicm}, the marginal distribution $(\ref{eq:3.marg})$ has the analytical form of a multivariate normal. The marginal distribution of $\ve{x}$ is a normal distribution $\ve x \sim N(\ve{0}, \ve{\Psi})$, with covariance matrix $\ve{\Psi}$ given in \eqref{Psi}. Hence, the marginal log-likelihood of $\ve{\theta}$ is given by
\begin{equation}
 l(\ve{\theta}|\D) = -\frac{1}{2} \log
|\ve{\Psi}(\ve{\theta})| - \frac{1}{2} \vesup{x}{\top}
\Psi(\ve{\theta})^{-1} \ve{x} - \frac{n}{2} \log 2 \pi.
\label{eq:3.lik1} \end{equation}
Thus, $\ve \theta$ is estimated by maximizing the above log-likelihood. 
The noise variance $\sigma_\epsilon^2$ can be estimated at the same time and we will henceforth treat it as one of the elements in $\ve \theta$.

The first and second derivatives of the log-likelihood \eqref{eq:3.lik1} with respect to the hyper-parameters of the covariance functions in Table~\ref{tab:statcovfuns} can be seen in Appendix~\ref{app:derivatives}.

\goodbreak

\subsection[Gaussian process regression with nonseparable and/or nonstationary covariance structure (NSGPR)]{Gaussian process regression with nonseparable and/or nonstationary covariance structure (NSGPR)} \label{subsec:NSGP}

A general class for nonstationary covariance functions \citep{konzen2020modelling} is given by
\begin{equation}\label{NScovfun}
\COV \big( f(\ve t),f(\ve t') \big) = \sigma(\ve t) \sigma(\ve t') | \ve A(\ve t) | ^{-1/4} | \ve A(\ve t') | ^{-1/4} \times \bigg| \frac{\ve A^{-1}(\ve t) + \ve  A^{-1}(\ve t')}{2} \bigg| ^{-1/2} g\Big( \sqrt{Q_{\ve t\ve t'}}\Big),
\end{equation}
where $g(\cdot)$ is a valid isotropic correlation function and
\begin{equation}\label{Qdist}
Q_{\ve t\ve t'} = (\ve t - \ve t')^\top \bigg( \frac{\ve  A^{-1}(\ve t) + \ve  A^{-1}(\ve t')}{2} \bigg) ^{-1} (\ve t - \ve t').
\end{equation}
The GPR model with nonseparable and/or nonstationary covariance function \eqref{NScovfun} will be referred to as NSGPR.

\cite{konzen2020modelling} propose to use B-spline basis functions for modeling the time or spatially varying unconstrained parameters in \eqref{NScovfun}, with spherical parametrization being used for the varying anisotropy matrix $\ve A(\ve t)$. These unconstrained parameters can be interpreted, via closed-form expressions, in terms of decay parameters and directions of dependence between the inputs. \cite{konzen2020modelling} show that the NSGPR model 
can extract important information from data with complex covariance structure by 
using a low-dimensional representation, which is based on the leading eigenvalues and 
eigensurfaces calculated from the estimated covariance structure.

\subsection[Multivariate Gaussian process regression (MGPR)]{Multivariate Gaussian process regression (MGPR)} \label{subsec:MGP}

Consider a multivariate function-valued process with dimension $p$, that is, ${\ve x(\ve t) = \left\{ x_1(\ve t), \dots, x_p(\ve t) \right\}^\top}$ defined on ${\cal T} \in \R^Q, \ Q \geq 1$. If we assume that $\ve x(\ve t)$ is a multivariate Gaussian function-valued process, then it is fully specified by its mean function $\ve \mu(\ve t) = \E ( \ve x(\ve t) )$ and covariance matrix function 
\[
\ve \Psi(\ve t, \ve t')  = \COV ( \ve x(\ve t), \ve x(\ve t') ) =  \left\{ \ve \Psi_{ij} (\ve t, \ve t') \right\}_{i,j=1}^p,
\]
where $\ve \Psi_{ij} (\ve t, \ve t') = \COV ( x_i(\ve t) , x_j(\ve t') )  $ denotes the auto-covariance functions (for $i=j$) and cross-covariance functions (for $i \neq j$).

A major difficulty is to construct flexible cross-covariance functions which are also valid: the mapping $\ve \Psi: \R^Q \times \R^Q \rightarrow M_{p \times p}$ must yield nonnegative definite matrices $M_{p \times p}$. In other words, the covariance matrix of the random vector $( \ve x(\ve t_1)^\top, \dots,  \ve x(\ve t_n)^\top )^\top \in \R^{np}$, given by
\begin{equation}\label{eq:SigmaMat}
\ve \Psi = 
\begin{bmatrix}
\ve \Psi(\ve t_1, \ve t_1) & \ve \Psi(\ve t_1, \ve t_2) & \cdots &  \ve \Psi(\ve t_1, \ve t_n) \\
\ve \Psi(\ve t_2, \ve t_1) & \ve \Psi(\ve t_2, \ve t_2) & \cdots &  \ve \Psi(\ve t_2, \ve t_n) \\
\vdots  & \vdots  & \ddots &  \vdots\\
\ve \Psi(\ve t_n, \ve t_1) & \ve \Psi(\ve t_n, \ve t_2) & \cdots &  \ve \Psi(\ve t_n, \ve t_n) 
\end{bmatrix},
\end{equation}
must be nonnegative definite.

Instead of defining covariance functions directly, one can use convolution methods to build valid covariance functions. For example, a stationary covariance function can be obtained by taking
\[
\ve \Psi_{ij}(\ve t, \ve t') = \int_{\R^Q}  k_i(\ve t - \ve t' - \ve u)k_j(\ve u)   d \ve u,
\]
where $k_i$ are square integrable functions satisfying $\sup \int_{\R^Q}  k_i^2(\ve u)   d \ve u < \infty$. Satisfying this condition is easier than specifying a covariance function directly, and different convolution specifications can be used to construct flexible covariance functions (e.g., \cite{higdon2002space}).

Based on these convolution processes, \cite{boyle2004dependent} suggest a bivariate convolution-based model by considering GPs constructed via convolution. This model is further discussed in \cite{shi2011gaussian}. Estimation of the covariance function hyper-parameters is conducted by using the concatenated observed responses and the corresponding covariance matrix \eqref{eq:SigmaMat} in the log-likelihood function. The \pkg{GPFDA} package uses a straightforward extension of this model to the case involving $p \geq 2$ outputs.

\goodbreak

\subsection[Gaussian process functional regression (GPFR)]{Gaussian process functional regression (GPFR)} \label{subsec:GPFR}

This section follows closely the exposition of \cite[Chapter 5]{shi2011gaussian}. Suppose we have a functional response variable $y_m(t)$, for $m=1, 2, \ldots, M$, a set of functional covariates $\vesub{x}{m}(t)$, and a set of scalar covariates $\vesub{u}{m}$, where
\[\vesub{x}{m}(t)=(x_{m1}(t), x_{m2}(t), \ldots, x_{mQ}(t))^\top \ \mbox{ and } \ \vesub{u}{m}=(u_{m1}, u_{m2}, \ldots, u_{mp})^\top. \]
 A general nonlinear regression model for the $m$th replication (curve, batch) is defined by
\begin{equation}
y_m(t)=f(t, \vesub{x}{m}(t), \vesub{u}{m}) + \epsilon_m(t),
\label{eq:5.m0}
\end{equation}
where $\epsilon_m(t)$'s are random errors which are independent at different $t$'s. 

A Gaussian process functional regression (GPFR) model is defined by
\begin{equation}
y_m(t) = \mu_m(t) + \tau_m(\vesub{x}{m}(t)) + \epsilon_m(t),
\label{eq:5.gpfr}
\end{equation}
where $\mu_m(t)$ is the common mean structure across different curves and $\tau_m(\vesub{x}{m}(t))$ defines the covariance structure of $y_m(t)$ for the different data points within the same curve. We use a GPR model to define the covariance structure:
\begin{equation}
\tau_m(\vesub{x}{m}(t)) \sim GPR_m[0, k_m(\vesub{\theta}{m})|\vesub{x}{m}(t)], \ \ m=1, \ldots, M,
\label{eq:5.gpfrcov}
\end{equation}
where $GPR_m[0, k_m(\vesub{\theta}{m})|\vesub{x}{m}(t)]$ denotes a GPR model with covariance function $k_m$ and hyper-parameters $\vesub{\theta}{m}$. Equations~\eqref{eq:5.gpfr} and \eqref{eq:5.gpfrcov} jointly define a GPFR model \citep{shi2011gaussian}, denoted by 
\[
y_m(t) \sim GPFR[\mu_m(t), k_m(\vesub{\theta}{m})|\vesub{x}{m}(t), \vesub{u}{m}].
\]

\subsubsection{GPFR model with a linear functional mean model} 

A special case of GPFR model is the case of a linear functional mean model. In particular, the mean model $\mu_m(t)$ is assumed to depend on scalar covariates $\vesub{u}{m}$ and $t$ only, through the linear relationship \mbox{$\mu_m(t)=\vess{u}{m}{\top} \ve{\beta}(t)$}. This special case is therefore given by
\begin{equation}
{y}_m(t)= \vess{u}{m}{\top} \ve{\beta}(t) + \tau_m(\vesub{x}{m}(t)) + \epsilon_m(t).
\label{eq:5.gpfrlin}
\end{equation}

Suppose that all the functional variables in the same batch are observed at the same data points $\{t_{mi}, i=1, \ldots, n_m\}$ for \mbox{$m=1, \ldots, M$}, so that the data observed in each batch are
\begin{equation}
{\cal D}_m =\{(y_{mi}, t_{mi}, x_{m1i}, \ldots, x_{mQi}), \mbox{ for } i =1, \ldots, n_m; \mbox{ and } (u_{m1}, \ldots, u_{mp}) \},
\label{eq:5.dm} \end{equation}
where  $y_{mi}=y(t_{mi})$ is the observation of $y_m(t)$ at $t_{mi}$ and  $x_{mqi}=x_q(t_{mi})$ is the measurement of the $q$th input variable for $q=1, \ldots, Q$. The observations of the scalar covariates for the $m$th batch are denoted by \mbox{$\vesub{u}{m}=(u_{m1}, \ldots, u_{mp})^\top$}. 


Both $y_m(t)$ and $\ve{\beta}(t)$ in \eqref{eq:5.gpfrlin} can be approximated by basis function representation:
\[
y_m(t) \approx \tilde{y}_m(t)=\vess{A}{m}{\top} \ve{\Phi}(t), \ \mbox{ and } \ \ve{\beta}(t) \approx \vesup{B}{\top} \ve{\Phi}(t), 
\]
where $\ve{\Phi}(t)=({\phi}_{1}(t), \ldots, {\phi}_{H}(t))^\top$ is a set of $H$ basis functions, $\vesub{A}{m}$ is an $H$-dimensional coefficient vector and  $\ve{B}$ is a $H \times p$ matrix.  For example, $\ve{\Phi}(t)$ could be a set of B-spline or Fourier basis functions.

Based on the data $\{ {\cal D}_m \}$ given in \eqref{eq:5.dm}, we can evaluate the marginal likelihood for the model (\ref{eq:5.gpfrlin}) and then calculate all unknown parameter estimates using an empirical Bayesian approach \citep[see][]{shi2007batch,shi2011gaussian}. 
In practice, we may use a fast approximation approach. The coefficients $\vesub{A}{m}$ and  $\ve{B}$ can be estimated by 
\begin{eqnarray*}
\hvesub{A}{m} & = & (\vesub{\Phi}{m}^\top \vesub{\Phi}{m})^{-1}\vesub{\Phi}{m}^\top \vesub{y}{m},\\
\hvesup{B}{\top} & = &(\vesup{U}{\top} \ve{U})^{-1} \vesup{U}{\top} \ve{A},
\label{eq:5.ham1} \end{eqnarray*}
where $\vesub{y}{m}=(y_{m1}, \ldots, y_{mn_m})^\top$, $\ve{U}=(\vesub{u}{1}, \ldots, \vesub{u}{M})^\top$ and $\vesub{\Phi}{m}$ is an $n_m \times H$ matrix with elements $(\phi_h(t_{mi}))$. The details can be found in Section 5.3 in \cite{shi2011gaussian}. 

\subsubsection{Predictions}

We now discuss how to calculate the prediction $y^*=y(t^*)$ at a new point $(t^*, \vesup{x}{*}, \vesup{u}{*})$ with $\vesup{x}{*} = \ve{x}(t^*)$. From (\ref{eq:5.gpfrlin}), the mean is estimated by
\begin{equation}
\hat{\mu}(t) = \vesup{u}{\top} \hvesup{B}{\top} \ve{\Phi}(t).
\label{eq:5.hatmu}
\end{equation}
The prediction of $y^*$ is given by
\begin{equation}
\hat{y}^* = \hat{\mu}(t^*) + \hat{\tau}(\vesup{x}{*}),
\label{eq:5.yhat}
\end{equation}
 where $\tau^*=\tau(\vesup{x}{*})$ is predicted by its conditional mean $\E(\tau^*|\D)$ from the GPR model defined in \eqref{eq:5.gpfrcov}. 

\uf{Type I prediction.} In addition to the training data which contains $M$ replications, suppose we now have also observed data for the $(M+1)$th replication and want to predict $y$ at a new data point $t^*$. Assume that $n$ observations have also been obtained in the new curve at $\ve{t}=(t_1, t_2, \ldots, t_n)^\top$, providing the data
\[
\D_{M+1}=\{(y_{M+1, i}, t_{M+1, i},x_{M+1,1, i}, \ldots, x_{M+1,Q, i}), \ i=1, \ldots, n; \ \vesub{u}{M+1} \}.
\]
Thus, the training data for prediction is $\D=\{ \D_1, \ldots, \D_M, \D_{M+1}\}$. 
To predict $y^*$ at a new  data point $t^*$, we assume that $y^*$ and the observed data ${\{y_{M+1, i}, \ i=1, \ldots, n\}}$ are generated from the same model (\ref{eq:5.gpfrlin}), and thus $\tau^*$ and ${\{\tau_{mi}, i=1, \ldots, n\}}$ have the same GPR model structure.  

The predictive mean $\hat{y}^*$ and predictive variance $\hat{\sigma}^{*2}$ are respectively given by Equations (5.26) and (5.27) in Section 5.3.1 of \cite{shi2011gaussian}. 

\uf{Type II prediction.} Now, suppose we have not observed any data besides the $M$ replications and want to make prediction for a completely new curve. 

We will keep the same notation, referring to the new curve as the $(M+1)$th curve and corresponding scalar covariates $\vesub{u}{M+1}$. Our objective is to predict $y^*$ at $(t^*, \vesup{x}{*})$ in the $(M+1)$th batch. In this case, there is no data observed in the $(M+1)$th batch, and thus the training data is $\D=\{ \D_1, \ldots, \D_M\}$. One simple method is to predict it using the mean part only, so that
\begin{equation}
\hat{y}^*=\hat{\mu}_{M+1}(t^*) = \vess{u}{M+1}{\top} \hvesup{B}{\top} \ve{\Phi}(t^*).
\label{eq:5.pmean2}
\end{equation}
Alternatively, we assume that curves $1, 2, \ldots, M$ provide an empirical distribution of the set of all possible curves \citep{shi2005hierarchical}, considering that
\begin{equation}
P(y^* \mbox{ belongs to the $m$th curve})=w_{m},
\label{eq:5.ptau}
\end{equation}
for $m=1, 2, \ldots, M$. 

Assuming that $y^*$ is generated from the $m$th curve means that the predictive mean and variance of $y^*$ can be calculated from the Type I prediction procedure. Therefore, a prediction for the response associated with a new input $\vesup{x}{*}$ at $t^*$ in a completely new curve can be calculated by
\begin{equation}
 \hat{y}^*=\sum_{m=1}^M w_m \hat{y}_m^*,
\label{newpmean}
\end{equation}
and the related predictive variance is
\begin{equation}
\hat{\sigma}^{*2}=\sum_{m=1}^M w_m \hat{\sigma}_m^{*2} +\left ( \sum_{m=1}^M w_m \hat{y}_m^{*2} - \hat{y}^{*2}\right ).
\label{newpvar}
\end{equation}
We usually take equal empirical probabilities, i.e., $w_m=1/M$. Unequal weights can be considered using an allocation model \cite[see, e.g.,][]{shi2008curve}.

\section[GPFDA package]{\pkg{GPFDA} package} \label{sec:package}

The main functions of the package can be seen in Table~\ref{tab:summary}. Each of the next subsections describes the functions used for estimation, prediction and visualization for each of the models discussed in Section~\ref{sec:meth}.

\begin{table}[H]
\centering
\begin{tabular}{lp{11cm}}
\hline
Function           & Description  \\ \hline
\code{cov.linear} &	Linear covariance function \\
\code{cov.matern} &	Stationary \Matern{} covariance function \\
\code{cov.pow.ex} &	Stationary powered exponential covariance function \\
\code{cov.rat.qu} &	Stationary rational quadratic covariance function \\
\code{gpfr} &	Gaussian process functional regression (GPFR) model  \\
\code{gpfrPredict} &	Prediction of GPFR model \\
\code{gpr} &	Gaussian process regression (GPR) model \\
\code{gprPredict} &	Prediction of GPR model \\
\code{mgpr} &	Multivariate Gaussian process regression (MGPR) model \\
\code{mgprPredict} &	Prediction of MGPR model \\
\code{nsgpr} &	Estimation of a nonseparable and/or nonstationary covariance structure (NSGPR model)\\
\code{nsgprPredict} &	Prediction of NSGPR model\\
\code{plot.gpfr} &	Plot GPFR model for either training or prediction \\
\code{plot.gpr} &	Plot GPR model for either training or prediction \\
\code{plot.mgpr} & Plot predictions of GPR model \\ \hline
\end{tabular}
\caption{\label{tab:summary} Summary of \pkg{GPFDA} package functions.}
\end{table}

\subsection[GPR]{GPR} \label{subsec:pkgGPR}

The function \fct{gpr} performs estimation of the GPR model \eqref{eq:2.basicm}. 
Its main arguments are

\begin{Code}
gpr(response, input, Cov = 'pow.ex', m = NULL, meanModel = 0, mu = NULL, 
  gamma = 2, nu = 1.5, useGradient = T, ...) 
\end{Code}

The input covariates and the response variable should be entered via the arguments 
\code{input} and \code{response}, respectively. 
Note that \code{response} can include multiple realizations.

The user may specify one or multiple covariance kernels (among \code{"linear"}, 
\code{"matern"}, \code{"pow.ex"}, and \code{"rat.qu"}) in a character vector 
passed to \code{Cov}. If multiple covariance kernels are informed, e.g., 
\code{Cov = c("matern", "linear")}, then the covariance function used for the 
GPR model will be the sum of these covariance kernels, each one applied to all 
input dimensions. The arguments \code{gamma} and \code{nu} are parameters for the \code{"pow.ex"} and \code{"matern"} classes.

If computational cost is a concern, the user may employ the Subset of Data method 
by choosing a sample size $m<n$, where $n$ is the sample size of each 
realization. These $m$ datapoints are randomly selected and reduce the time 
complexity of the GPR model estimation from ${\cal O} (n^3)$ to ${\cal O} (m^3)$.

By means of the argument \code{meanModel}, the user can choose one of the following mean function models: 
zero mean function, constant, linear model, or the average across replications 
(provided the multiple realizations are observed at the same covariate values).
Alternatively, the user can specify the values for the mean function directly in the argument \code{mu}.

Optimization is performed using the function \fct{nlminb} of the \pkg{stats} package.  If \code{useGradient = TRUE} is specified, analytical expressions for gradients are used. Note that for the \Matern{} covariance class the gradient is only available for the cases $\nu=3/2$ and $\nu=5/2$. The noise variance $\sigma_\epsilon^2$ and the hyper-parameters are estimated at the same time, since \fct{gpr} treats $\sigma_\epsilon^2$ as one of the elements in $\ve \theta$.

\fct{gpr} returns an object of class \class{gpr} containing many results from the estimated GPR model, including the estimated hyper-parameters and mean function. The \class{gpr} class object can be used directly in the argument \code{train} in the function \fct{gprPredict}, which provides predictions for every input $\ve t^*$ in \code{inputNew}:

\begin{Code}
gprPredict(train = NULL, inputNew = NULL, noiseFreePred = F, mSR = NULL, ...)
\end{Code}

The Subset of Regressors approximation method for predictions can be used by entering 
an integer value in \code{mSR}; in this case, a subset of \code{mSR} columns will be randomly chosen.

The user can choose to see predictions which are noise-free or not by means of the 
argument \code{noiseFreePred}. If \code{noiseFreePred = TRUE} is specified, noise-free predictions are obtained by setting 
$\sigma_\epsilon^2=0$ in \eqref{eq:2.pr30}.

If no object is provided to \code{train}, learning is conducted based on the other arguments of \fct{gprPredict}. \fct{gprPredict} returns an object which includes the mean and standard deviation of predictions (\code{pred.mean} and \code{pred.sd}). This object can be used directly in \fct{plot} method to visualize predictions with confidence intervals.

\subsection[NSGPR]{NSGPR} \label{subsec:pkgNSGPR}

Analogously, for the NSGPR model with covariance function \eqref{NScovfun}, \code{input} and \code{response} are required arguments in the function \fct{nsgpr}:

\begin{Code}
nsgpr(response, input, corrModel = "pow.ex", gamma = 2, nu = 1.5, 
  whichTau = NULL, nBasis = 5, cyclic = NULL, unitSignalVariance = F, 
  zeroNoiseVariance = F, sepCov = F, ...)
\end{Code}

In \code{corrModel}, the user specifies the correlation function model for $g(\cdot)$ 
in \eqref{NScovfun} which can be can be either \code{"pow.ex"} or \code{"matern"}.

For multidimensional inputs, the argument \code{whichTau} identifies which input coordinates the parameters are function of. The argument \code{cyclic} defines which covariates are cyclic
(periodic). \code{nBasis} is the number of B-spline basis functions to be used.

If the function-valued process is known to have unit variance, 
\code{unitSignalVariance} can be set to \code{TRUE}. Similarly, 
\code{zeroNoiseVariance} should be \code{TRUE} if realizations are 
assumed to be noise-free. The argument \code{sepCov} controls whether 
off-diagonal elements of the varying anisotropy matrix should be set to zero. 

Maximum likelihood estimates of B-spline coefficients and noise variance are returned 
by \fct{nsgpr}. These hyper-parameter estimates can be used in the argument \code{hp} 
in \fct{nsgprPredict} in order to obtain predictions at new input $\ve t^*$.

\subsection[MGPR]{MGPR} \label{subsec:pkgMGPR}

To fit the multivariate GP model discussed in Section~\ref{subsec:MGP}, the 
following code is used:
\begin{Code}
mgpr(Data, m = NULL, meanModel = 0, mu = NULL)
\end{Code} 

\code{Data} should be a list including both input and response variables.  
The arguments \code{m}, \code{meanModel} and \code{mu} are used as in \fct{gpr}, 
with the mean function specifications applied to each response variable separately.

\fct{mgpr} returns an \class{mgpr} class object with the results of the 
estimated MGPR model. This object can then be passed to the argument \code{train} of the function \fct{mgprPredict} to obtain predictions. Finally, the \fct{plot} method can be used to visualize predictions given an \class{mgpr} object.


\subsection[GPFR]{GPFR} \label{subsec:pkgGPFR}

The function \fct{gpfr} performs estimation of \eqref{eq:5.gpfr}. Its main arguments are

\begin{Code}
gpfr(response, time = NULL, uReg = NULL, fxReg = NULL, fyList = NULL, 
  uCoefList = NULL, fxList = NULL, concurrent = TRUE, fxCoefList = NULL, 
  gpReg = NULL, Cov = "pow.ex", gamma = 2, nu = 1.5, fitting = F, ...)
\end{Code} 

For the mean function $\mu_m(t)$, a FR model is used and can include scalar covariates $\vesub{u}{m}$ and functional covariates $\vesub{x}{m}(t)$. For scalar covariates, a regression coefficient function $\ve{\beta}(t)$ is fitted for the linear functional mean model \mbox{$\mu_m(t)=\vess{u}{m}{\top} \ve{\beta}(t)$}. For functional covariates, a regression coefficient vector $\ve{\alpha}$ is estimated for the model \mbox{$\mu_m(t)=\ve{\alpha}^\top \vesub{x}{m}(t)$} (if \code{concurrent} is set to \code{FALSE}) or a regression coefficient function $\ve{\alpha}(t)$ is estimated for the functional concurrent model \mbox{$\mu_m(t)=\ve{\alpha}(t) \vesub{x}{m}(t)$} (if \code{concurrent} is set to \code{TRUE}).
The residual part of \eqref{eq:5.gpfr} is modeled by a GP $\tau_m$ with zero mean and covariance function depending on $t$ or functional covariates $\vesub{x}{m}(t)$.

As covariate(s) for the Gaussian process $\tau_m(\cdot)$, the user should enter the input $t$ or functional covariates $\vesub{x}{m}(t)$ in \code{gpReg}.

For the mean function $\mu_m(t)$, the user should enter data through the arguments \code{uReg} (scalar regressors) and \code{fxReg} (functional regressors). The function \fct{gpfr} knows what model to use by checking the informed arguments. The FR model will be estimated including 
(i)  scalar covariates $\vesub{u}{m}$ if some data are entered in \code{uReg}; 
and/or (ii) functional covariates $\vesub{x}{m}(t)$ if some data are passed to \code{fxReg}. 

To deal with the functional terms $y_m(t)$, $\ve{\beta}(t)$, $\ve{\alpha}(t)$, and $\vesub{x}{m}(t)$, the user can set up the functional variables with special options. These options are used to build ``\code{fd}'' objects through functionalities of the \code{fda} package. The performance of the fitted GPFR model \eqref{eq:5.gpfr} depends on the performance of the FR model used for the mean function. The key point is the amount of smoothness introduced into the functional part. Too much smoothness may result larger bias in the fitting and prediction results. For all functional variables or functional coefficients, the smoothness is primarily controlled by the number of basis functions and the value of tuning parameters of the roughness penalty. There are default specifications for all options, and they are intended to give a good answer for most of the cases. However, we explain below how the user can modify them if it is desired to do so.

Customization of the functional variable $y_m(t)$  can be done by setting up a list for \code{fyList} containing the following specifications:
\begin{itemize}
\item \code{time}: a sequence of time points for $t$ (default are 100 points from 0 to 1).
\item \code{nbasis}: number of basis functions used in smoothing (default is the minimum between one fifth of the time points and 23).
\item \code{norder}: number of basis functions used in smoothing (default is 6).
\item \code{bSpline}: logical. If \code{TRUE} (default), B-spline basis is used, if \code{FALSE}, Fourier basis is used.
\item \code{Pen}: penalty term in the smoothing. The default is \code{c(0, 0)}, meaning that the penalty is only applied to the second order derivative of the curve, with no penalty for the zero-th and first order derivatives of the curve; if Fourier basis is used, the default penalty will be based on the harmonic acceleration function $c(0, 1, 0, \omega^2)$, where $\omega$ is the period of the basis function.
\item \code{lambda}: smoothing parameter for the penalty (default $10^{-4}$).
\end{itemize}

Customization of the functional covariates $\vesub{x}{m}(t)$ can be done similarly by choosing specifications of \code{fxList}. The only difference is that \code{fxList} is a list of lists, in order to allow for different specifications for each functional covariate if there are multiple ones.

\code{uCoefList} and \code{fxCoefList} are similar to each other and are useful to choose the specifications of $\ve{\beta}(t)$ (functional coefficients of scalar covariates) and $\ve{\alpha}(t)$ (functional coefficients of functional covariates in the concurrent model). Each of them is also expected to be a list of lists. Their specifications \code{nbasis}, \code{norder}, \code{bSpline}, \code{Pen}, and \code{lambda} are similar as in \code{fyList}, having different default values. 

To visualize how the functional variables look like for a given customization, \fct{mat2fd} followed by \fct{plot} may be used.

Finally, \fct{gpfr} will return the in-sample fitted values with standard deviation if
\code{fitting} is set to \code{TRUE}.

Given a \class{gpfr} class object obtained by \fct{gpfr}, \fct{gpfrPredict} returns an object containing predictions with associated standard deviations. These will be Type I predictions if some data are passed to the argument \code{gpReg} or Type II predictions   otherwise. For calculating Type II predictions, equal empirical probabilities $w_m=1/M$ in \eqref{eq:5.ptau} are used. The resulting object obtained by \fct{gpfrPredict} can be passed to \fct{plot} for visualization of predictions.

\section{Examples}\label{sec:examples}

\pkg{GPFDA} package provides several examples illustrating each model in \textit{vignettes}. Vignettes ``\code{gpr_ex1}'' and ``\code{gpr_ex2}'' explain how to conduct estimation and prediction of GPR models with one- and two-dimensional covariates, respectively. Vignette ``\code{co2}'' shows, through an application to \code{co2} data, how the users can customize their own covariance kernel. Implementation of NSGPR models can be seen in vignette ``\code{nsgpr}''. In the next subsections we show examples for MGPR and GPFR models which are described in more details in the vignettes ``\code{mgpr}'' and ``\code{gpfr}'', respectively.

\subsection{MGPR example}

We simulate $30$ realizations from a trivariate process
${\ve x(t) = \left\{ x_1(t), x_2(t), x_3(t) \right\}^\top}$, where ${t \in \R}$, 
following the model discussed in Section~\ref{subsec:MGP}. Each response 
variable $x_j(t)$ is observed on $250$ equally spaced time points. 
More details can be seen in the package vignette ``\code{mgpr}''.

\begin{figure}[H]
  \begin{minipage}[b]{1.9\linewidth}
\begin{Schunk}
\begin{Sinput}
R> old <- par(mfrow = c(1,3), mar = c(4.5,5.1,0.2,0.8), oma = c(0,0,0,0))
R> for(j in 1:3){
+    matplot(Data$input[[j]], Data$response[[j]], type = "l", lty = 1, 
+      xlab = "t", ylab = bquote(x[.(j)]), cex.lab = 2, cex.axis = 1.5)
+  }
R> par(old)
\end{Sinput}
\end{Schunk}
\includegraphics{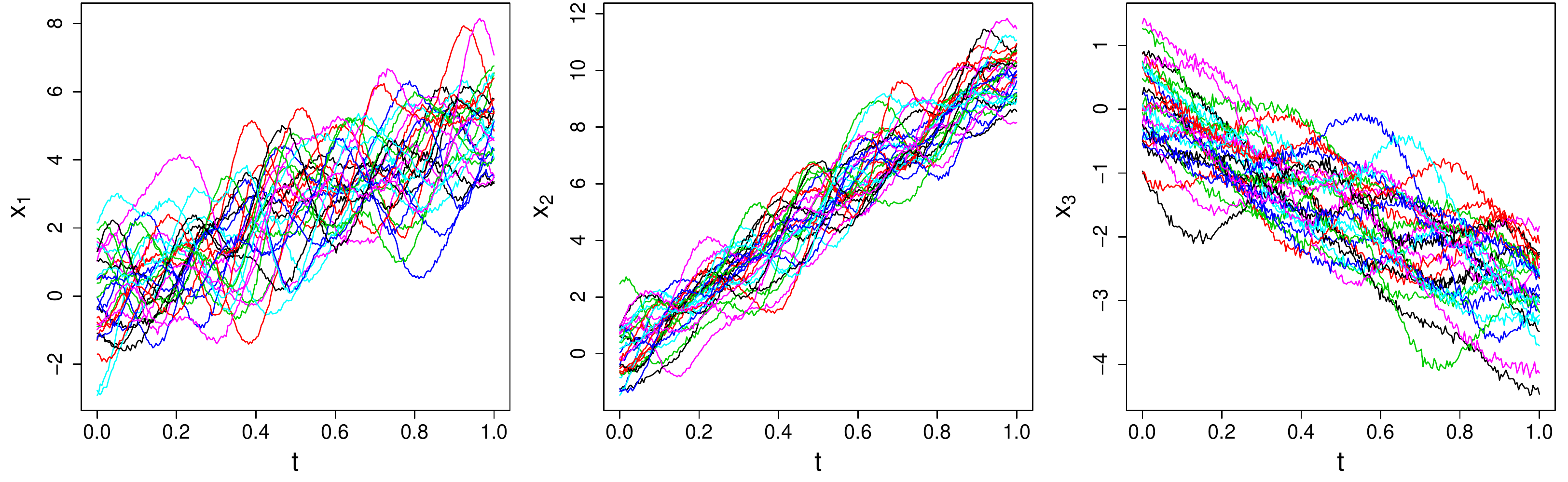}
  \end{minipage}%
\caption{Trivariate data used for the MGPR example.}\label{fig:mgprSimData}
\end{figure}

These data are saved in \pkg{GPFDA} package under the name \code{dataExampleMGPR}. After installing and loading \pkg{GPFDA}, these data can be loaded by using the following \proglang{R} command:
\begin{Schunk}
\begin{Sinput}
R> data("dataExampleMGPR")
R> Data <- dataExampleMGPR
\end{Sinput}
\end{Schunk}
The simulated data can be visualized in Figure~\ref{fig:mgprSimData}.
 
Suppose we want to estimate the MGPR model of Section~\ref{subsec:MGP} for these data assuming a linear function for each response variable and using a subset of $m=100$ randomly selected datapoints of each realization in the model estimation.
\begin{Schunk}
\begin{Sinput}
R> set.seed(123)
R> mgprFit <- mgpr(Data = Data, m = 100, meanModel = 't')
\end{Sinput}
\end{Schunk}
Based on the estimated model, suppose we want to predict the values of the three response variables at $60$ new time points. These time points will be stored in the object \code{DataNew}.
\begin{Schunk}
\begin{Sinput}
R> n_star <- 60
R> input1star <- input2star <- input3star <- seq(0, 1, length.out = n_star)
R> DataNew <- list()
R> DataNew$input <- list(input1star, input2star, input3star)
\end{Sinput}
\end{Schunk}

\begin{figure}[H]
  \begin{minipage}[b]{1.8\linewidth}
\begin{Schunk}
\begin{Sinput}
R> plot(mgprFit, DataObs = DataObs, DataNew = DataNew)
\end{Sinput}
\end{Schunk}
\includegraphics{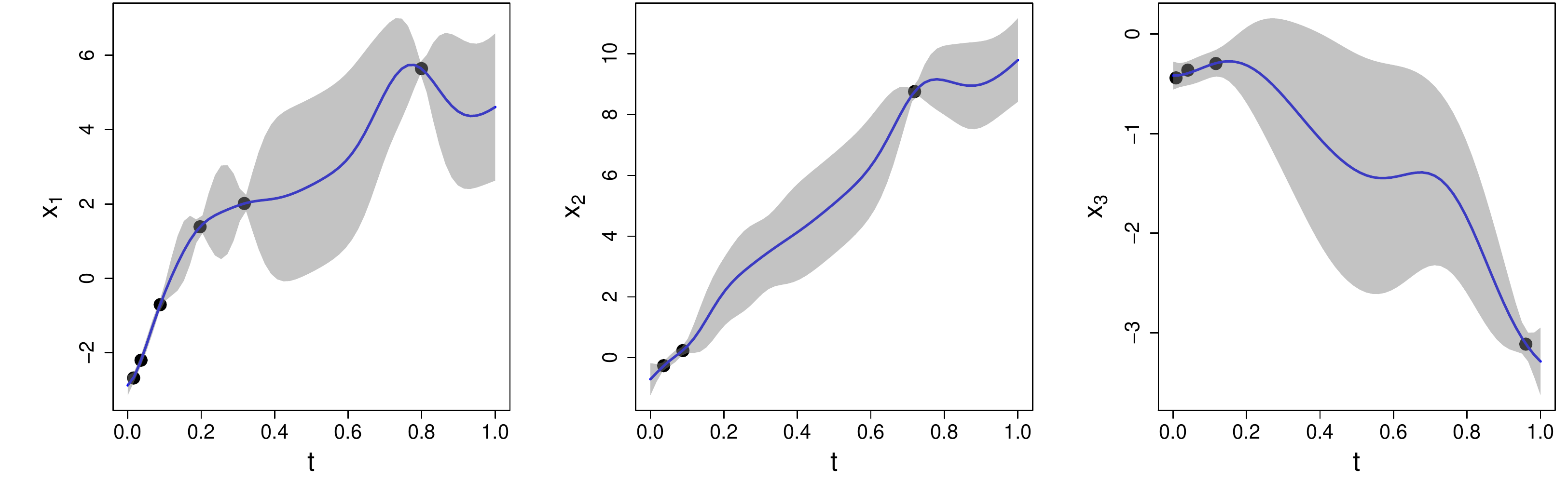}
  \end{minipage}%
\caption{MGPR predictions for the fifth trivariate curve given a small subset of datapoints.}
\label{fig:mgprPred1}
\end{figure}

We have trained the model using $m$ time points. However, for visualization purposes, suppose we want to see predictions based on very few data points.
We will use a very small subset of observations of the fifth 
realization of the simulated \code{Data} and save this subset in an object called \code{DataObs}. Based on this subset of very few observed time points, 
we will make predictions for the corresponding trivariate curve at the $60$ 
new time points. These predictions (with $95\%$ confidence interval) can be seen in Figure~\ref{fig:mgprPred1}.
\begin{Schunk}
\begin{Sinput}
R> idx <- 5
R> obs <- list()
R> obs[[1]] <- c(5, 10, 23, 50, 80, 200)
R> obs[[2]] <- c(10, 23, 180)
R> obs[[3]] <- c(3, 11, 30, 240)
R> DataObs <- list()
R> DataObs$input[[1]] <- Data$input[[1]][obs[[1]]]
R> DataObs$input[[2]] <- Data$input[[2]][obs[[2]]]
R> DataObs$input[[3]] <- Data$input[[3]][obs[[3]]]
R> DataObs$response[[1]] <- Data$response[[1]][obs[[1]], idx]
R> DataObs$response[[2]] <- Data$response[[2]][obs[[2]], idx]
R> DataObs$response[[3]] <- Data$response[[3]][obs[[3]], idx]
\end{Sinput}
\end{Schunk}

\begin{figure}[H]
  \begin{minipage}[b]{1.8\linewidth}
\begin{Schunk}
\begin{Sinput}
R> obs[[1]] <- c(5, 10, 23, 50, 80, 100, 150, 200)
R> obs[[2]] <- c(10, 23, 100, 150, 180)
R> DataObs$input[[1]] <- Data$input[[1]][obs[[1]]]
R> DataObs$input[[2]] <- Data$input[[2]][obs[[2]]]
R> DataObs$response[[1]] <- Data$response[[1]][obs[[1]], idx]
R> DataObs$response[[2]] <- Data$response[[2]][obs[[2]], idx]
R> plot(mgprFit, DataObs = DataObs, DataNew = DataNew)
\end{Sinput}
\end{Schunk}
\includegraphics{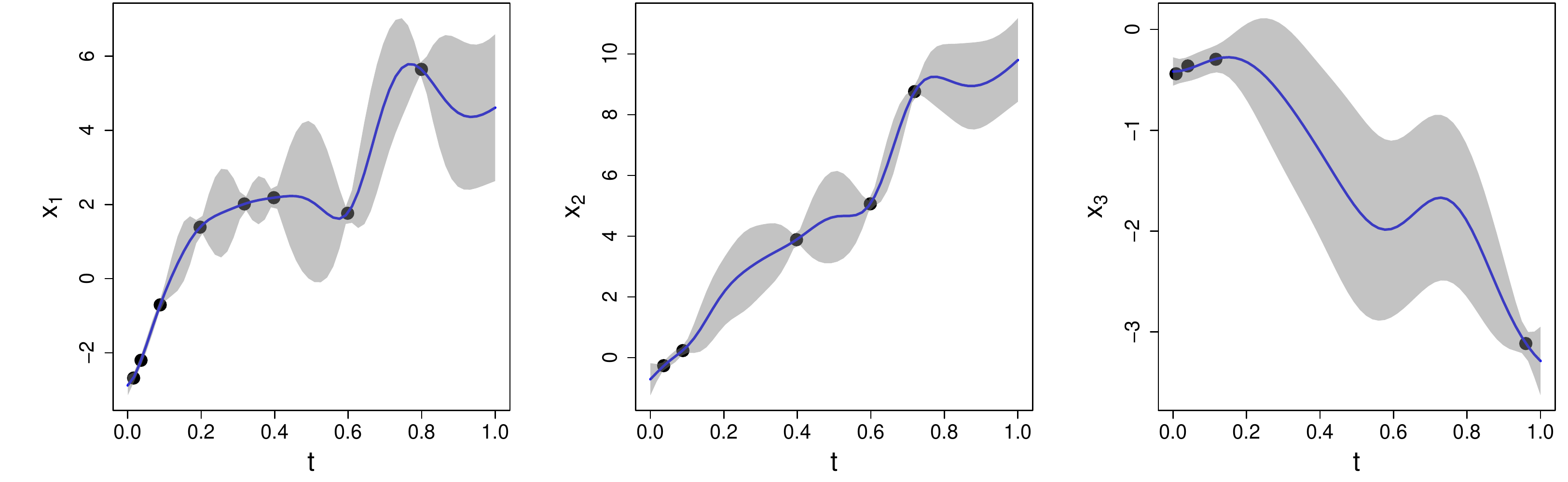}
  \end{minipage}%
\caption{MGPR predictions for the fifth trivariate curve given a larger subset of datapoints.}
\label{fig:mgprPred2}
\end{figure}

Suppose we have now observed two additional datapoints of the first two response variables -- their $100$th and $150$th observations in \code{Data}. Predictions based on these new observed data are shown in Figure~\ref{fig:mgprPred2}. We can notice how predictions for the first two response variables have now much less uncertainty in the middle of the time interval (compared to Figure~\ref{fig:mgprPred1}). In addition, predictions for the third response are affected by the information added to the other functions.


\goodbreak
 
\subsection{GPFR example}

Suppose we have a functional response variable $y_m(t), \ m=1,\dots,M$, a 
functional covariate $x_m(t)$ and a set of two scalar covariates 
$\textbf{u}_m = (u_{m0},u_{m1})^\top$. The GPFR model is therefore
\begin{equation}
y_m(t) = \mu_m(t) + \tau_m(x_m(t)) + \varepsilon_m(t),
\end{equation}
where $\mu_m(t) = \textbf{u}_m^\top \boldsymbol{\beta}(t)$ is the mean function
model across different curves and $\tau_m(x_m(t))$ is a GP with 
zero mean and covariance function $k_m(\boldsymbol{\theta}|x_m(t))$. That is, 
$\tau_m(x_m(t))$ defines the covariance structure of $y_m(t)$ for the different
data points within the same curve. The error term is assumed to be $\varepsilon_m(t) \sim N(0, \sigma_\varepsilon^2)$, where the noise variance $\sigma_\varepsilon^2$ can be estimated as a hyper-parameter of the GP.

In the example below, the training data consist of $M=20$ realizations on $[-4,4]$ with 
${n=50}$ points for each curve. We assume regression coefficient functions
${\beta_0(t)=1}$ and ${\beta_1(t)=\sin((0.5 t)^3)}$, scalar covariates 
$u_{m0} \sim N(0,1)$ and $u_{m1} \sim N(10,5^2)$, and a functional covariate
$x_m(t) = \exp(t) + v$, where $v \sim N(0, 0.1^2)$. The term $\tau_m(x_m(t))$ is
a zero mean GP with exponential covariance kernel and 
$\sigma_\varepsilon^2 = 1$.

Using the same data generating process, we simulate an independent $(M+1)$th 
realization which will be used to assess predictions obtained by the model 
estimated by using the training data of size $M$. The 
$y_{M+1}(t)$ and $x_{M+1}(t)$ curves are observed on equally spaced $60$ time points on $[-4,4]$.

The package vignette ``\code{gpfr}'' explains how these data were simulated. The package has this dataset saved under the name \code{dataExampleGPFR}, which can be loaded as follows:
\begin{Schunk}
\begin{Sinput}
R> data("dataExampleGPFR")
R> attach(dataExampleGPFR)
\end{Sinput}
\end{Schunk}
This loads several elements including the training data (with $M$ realizations) and the test data ($(M+1)$th realization). The estimation of the GPFR model is done by
\begin{Schunk}
\begin{Sinput}
R> gpfrFit <- gpfr(response = response_train, time = tt, uReg = scalar_train, 
+    gpReg = x_train, fyList = list(nbasis = 23, lambda = 0.0001), 
+    uCoefList = list(list(lambda = 0.0001, nbasi = 23)), 
+    Cov = 'pow.ex', gamma = 1, fitting = T)
\end{Sinput}
\end{Schunk}

\begin{figure}[H]
\centering
  \begin{minipage}[b]{0.9\linewidth}
\includegraphics{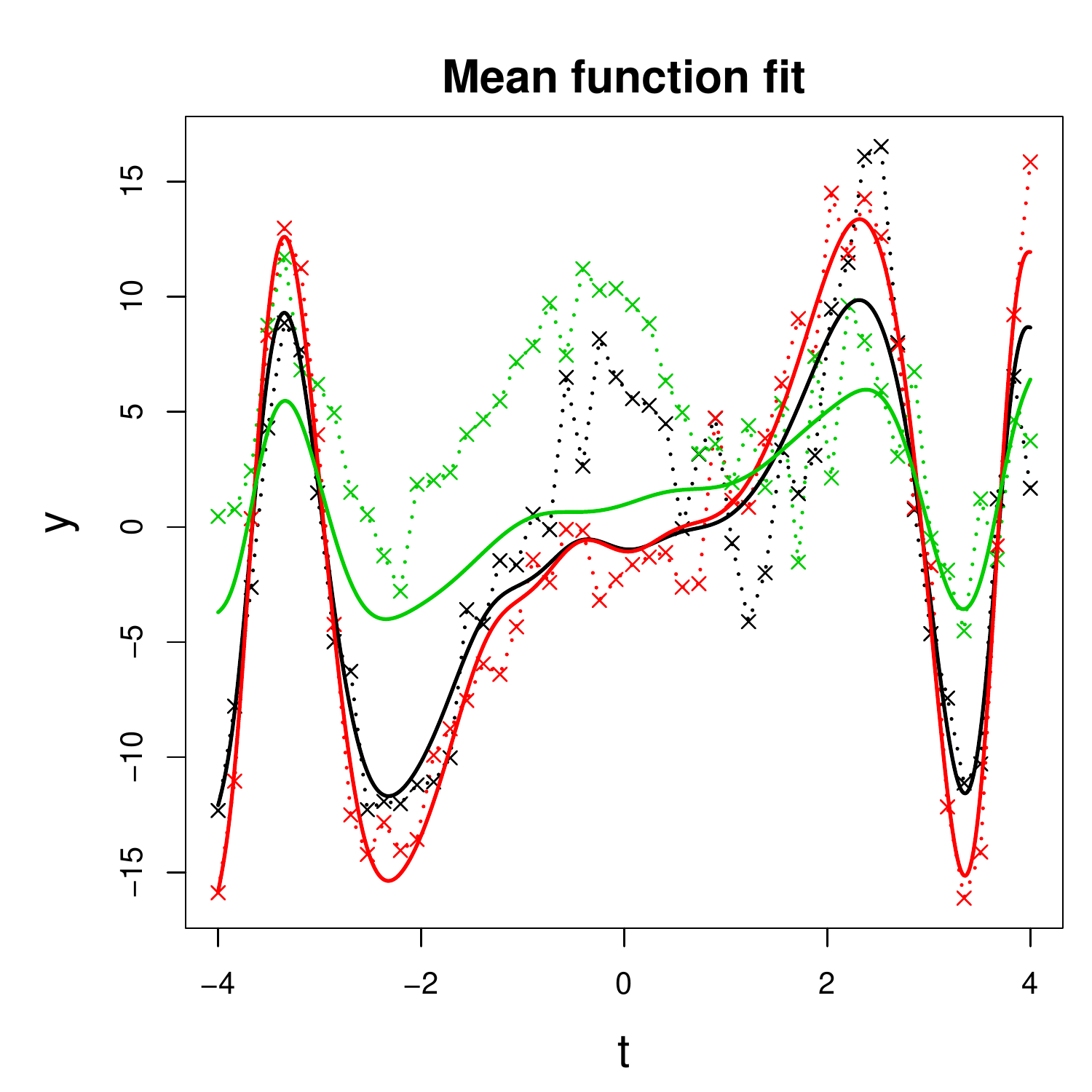}
  \end{minipage}%
  \begin{minipage}[b]{0.9\linewidth}
     \hspace{-7cm}
\includegraphics{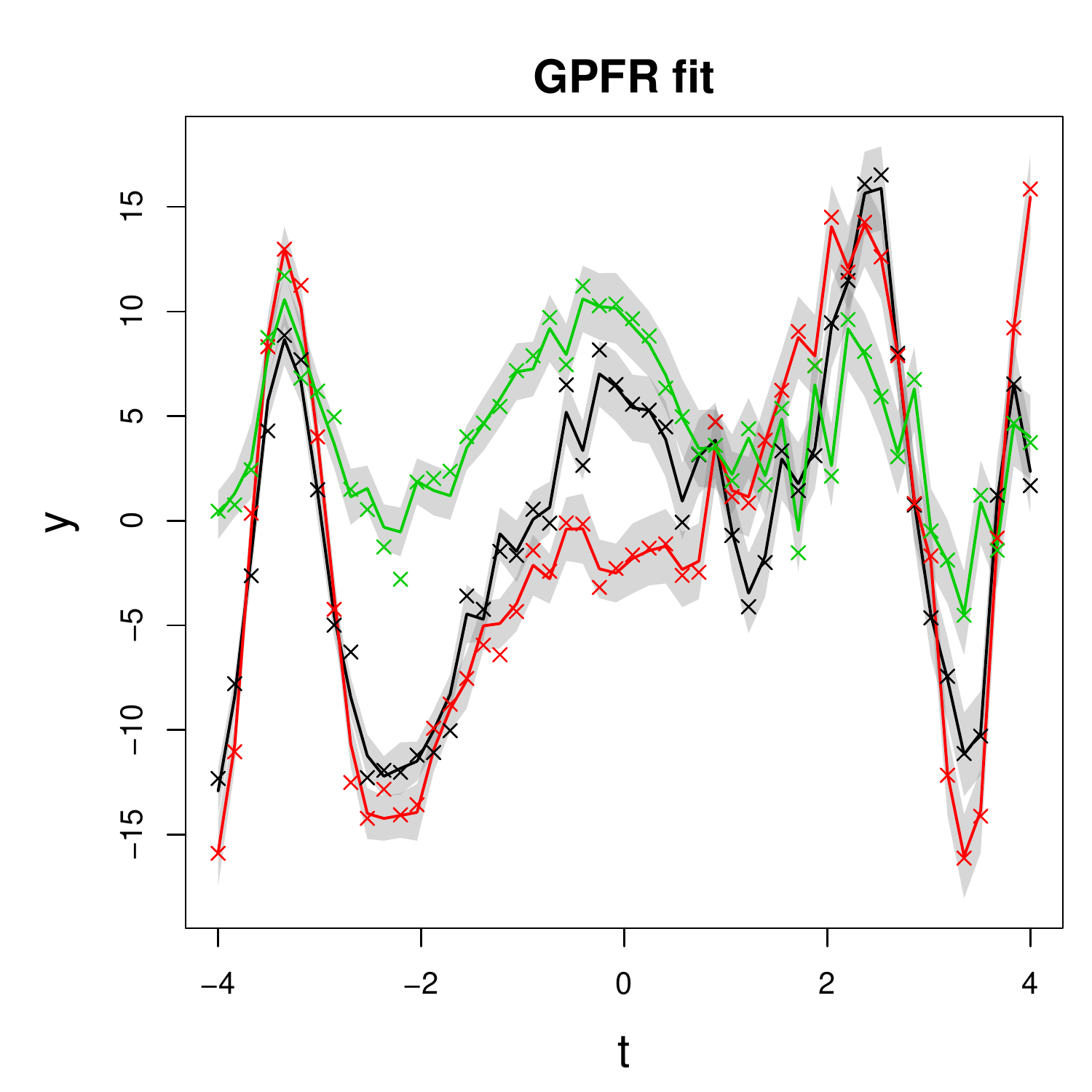}
  \end{minipage}
  \vspace{-0.5cm}
  \caption{Fitting results for the first three realizations of the sample. Left: mean function fit obtained by the FR model. Right: GPFR fit. Observed datapoints are represented by crosses and fitting results by solid lines. 95\% confidence intervals are shown in grey.}\label{fig:meanFRfittedGPFR}
\end{figure}

The mean function, which is estimated by the FR model, and the GPFR model fit for three realizations can be seen in Figure~\ref{fig:meanFRfittedGPFR}. These are obtained by
\begin{Schunk}
\begin{Sinput}
R> plot(gpfrFit, type = 'meanFunction', realisations = 1:3)
R> plot(gpfrFit, type = 'fitted', realisations = 1:3)
\end{Sinput}
\end{Schunk}

\begin{figure}[H]
 \centering
  \begin{minipage}[b]{0.9\linewidth}
%
\includegraphics{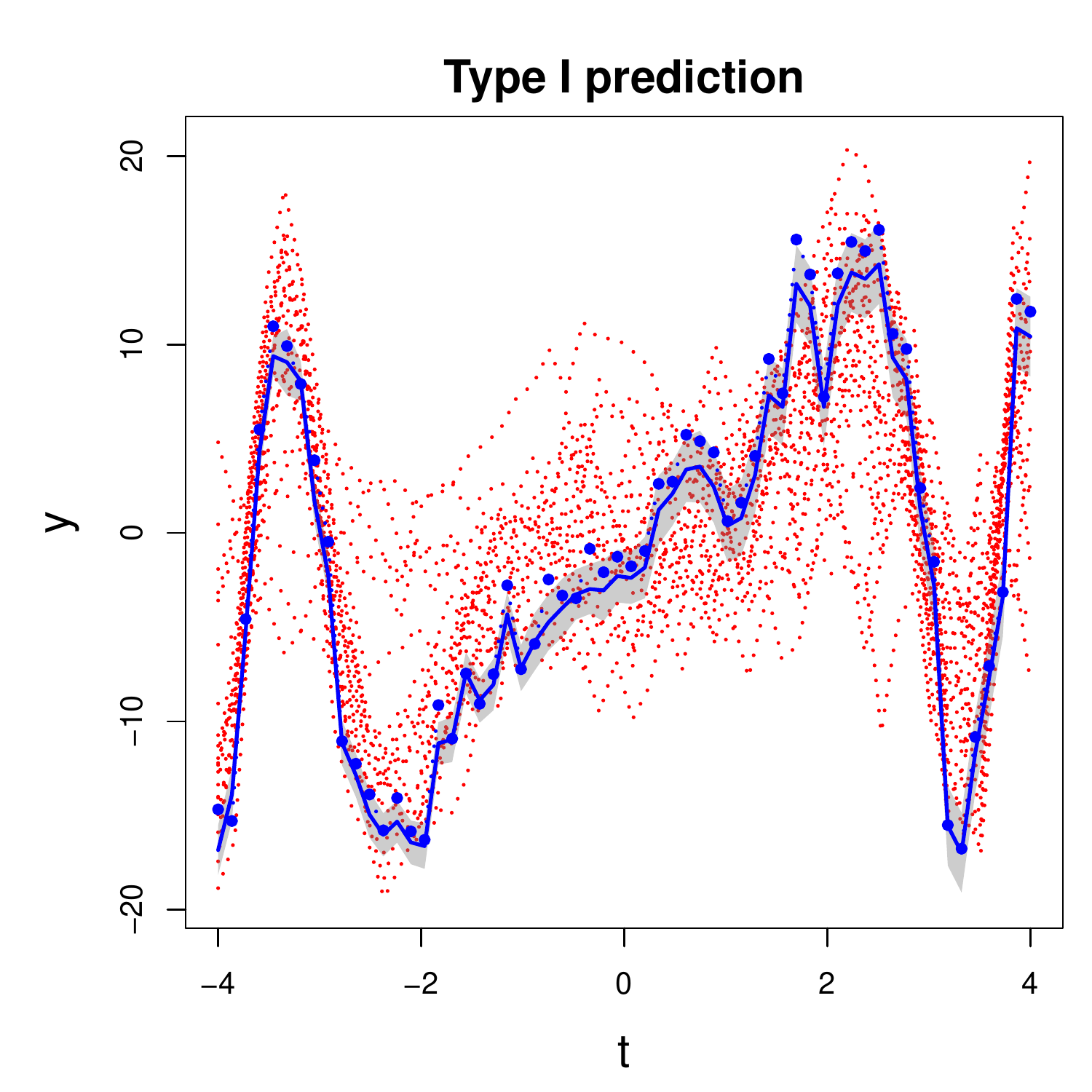}
  \end{minipage}%
  \begin{minipage}[b]{0.9\linewidth}
     \hspace{-7cm}
\includegraphics{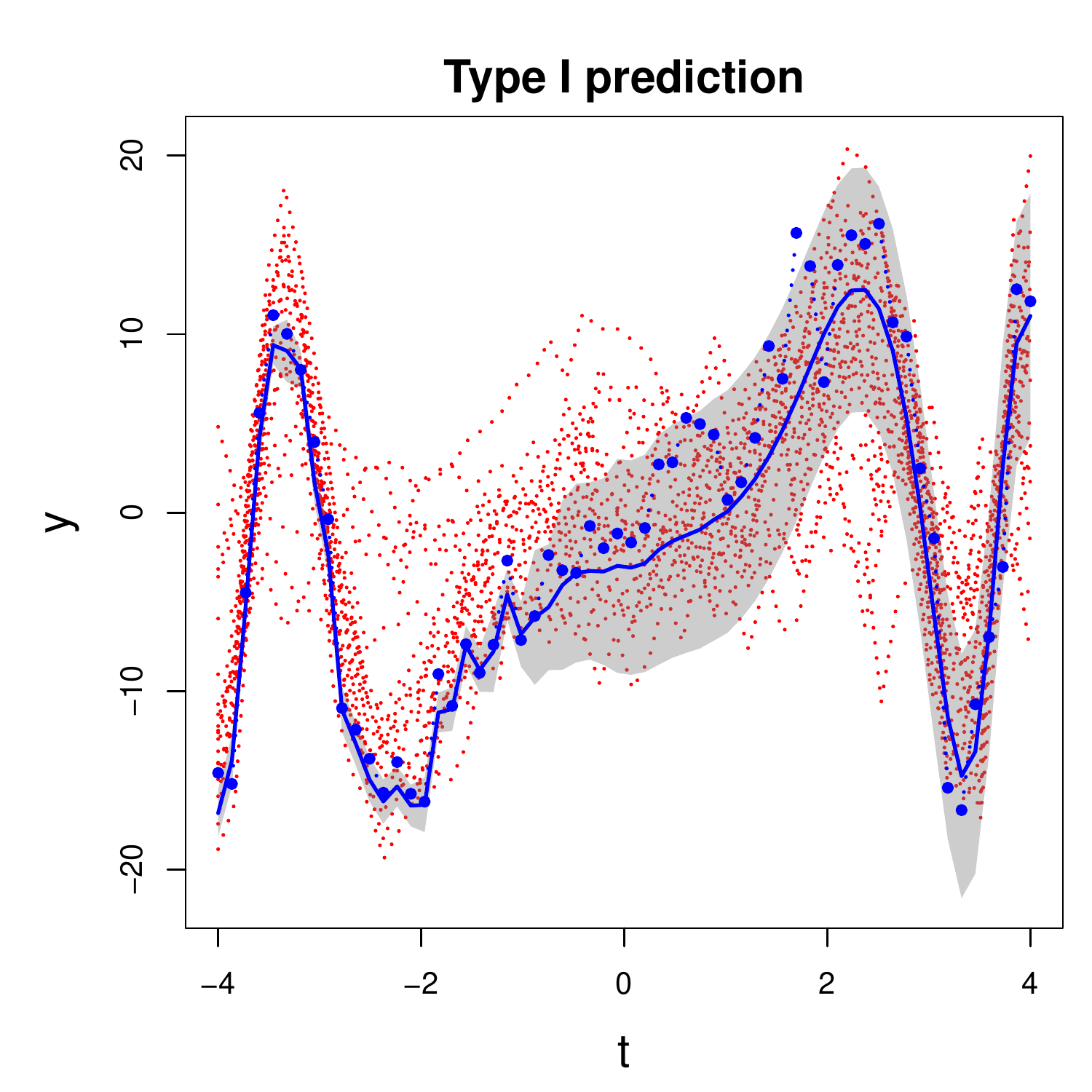}
    \end{minipage}%
  \caption{Type I predictions. The blue points represent the datapoints of $y_{M+1}(t)$ and the red dotted lines represent the other $M$ curves. The blue solid lines are the predictive means and the grey areas the corresponding $95\%$ confidence interval. Left: predictions use information from all the $60$ datapoints of the new curve which are observed over all the input domain. Right: predictions only take into account the first $20$ datapoints of the new curve.}
  \label{fig:typeIpred}
\end{figure}

Suppose we have the information about the $60$ datapoints of the new curve $y_{M+1}(t)$ which are observed over all the domain . Given these datapoints, Type I predictions can be seen on the left side of Figure~\ref{fig:typeIpred}. They are obtained as follows.
\begin{Schunk}
\begin{Sinput}
R> gpfrPredType1a <- gpfrPredict(train = gpfrFit, testInputGP = x_new, 
+    testTime = t_new, uReg = scalar_new, 
+    gpReg = list('response' = response_new, 'input' = x_new, 'time' = t_new))
R> plot(gpfrPredType1a, type = 'prediction')
R> lines(t_new, response_new, type = 'b', col = 4, pch = 19, cex = 0.6, 
+    lty = 3, lwd = 2)
\end{Sinput}
\end{Schunk}
If we now assume that $y_{M+1}(t)$ is only partially observed (using only the 
first one third of datapoints of $y_{M+1}(t)$), the Type I predictions can be obtained by
\begin{Schunk}
\begin{Sinput}
R> gpfrPredType1b <- gpfrPredict(train = gpfrFit, testInputGP = x_new, 
+    testTime = t_new, uReg = scalar_new, 
+    gpReg = list('response' = response_new[1:20], 
+    'input' = x_new[1:20], 'time' = t_new[1:20]))
R> plot(gpfrPredType1b, type = 'prediction')
R> lines(t_new, response_new, type = 'b', col = 4, pch = 19, cex = 0.6,
+    lty = 3, lwd = 2)
\end{Sinput}
\end{Schunk}
These new predictions are displayed on the right side of Figure~\ref{fig:typeIpred}. Note the larger uncertainty in the region of $t$ where we no longer use information about $y_{M+1}(t)$.

\begin{figure}[H]
\centering
  \begin{minipage}[b]{0.9\linewidth}
  \centering
\includegraphics{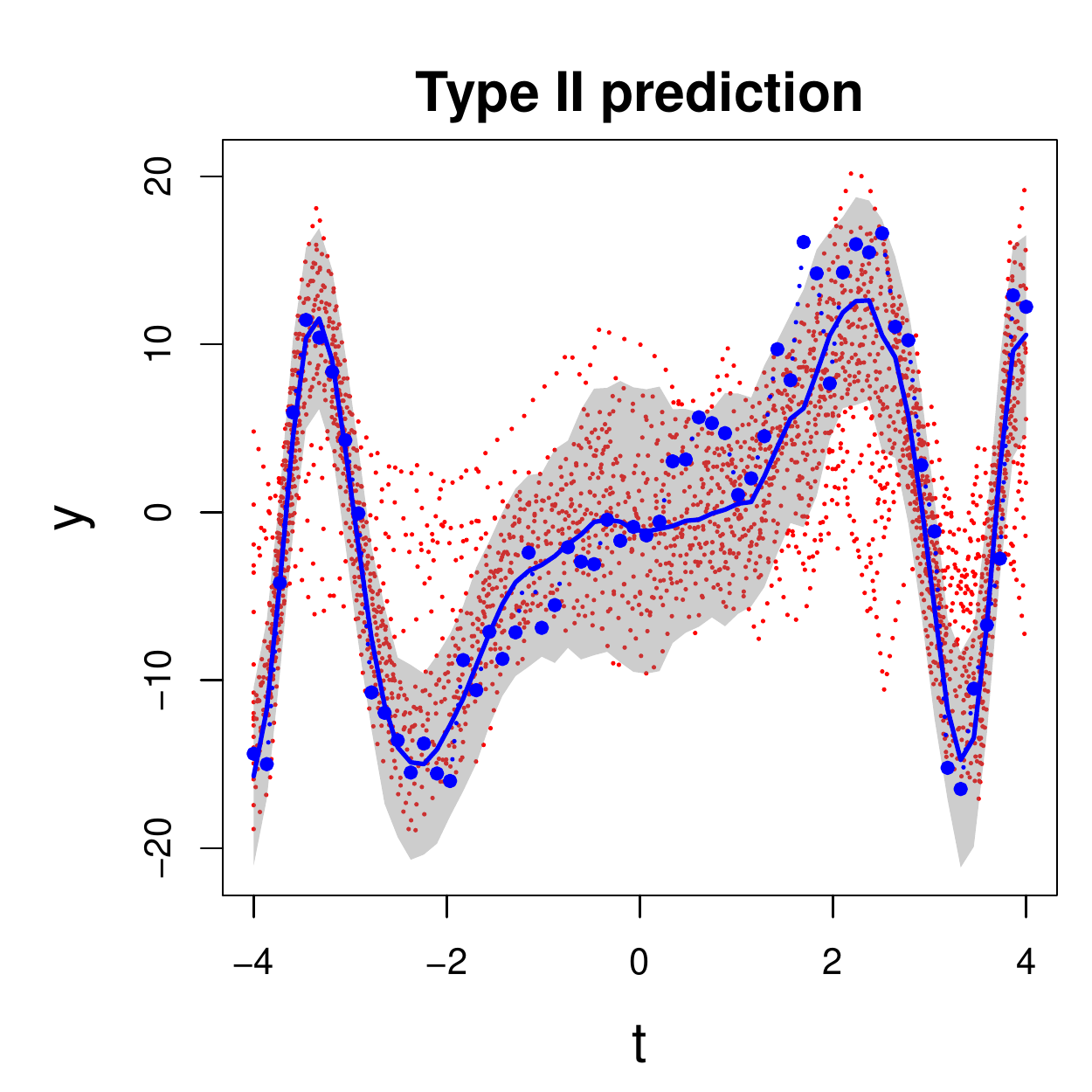}
    \end{minipage}%
\caption{Type II predictions, i.e, no information about $y_{M+1}(t)$ is used.}\label{fig:typeIIpred}
\end{figure}

Finally, the Type II prediction, which is made by not including any information about $y_{M+1}(t)$, is visualized in Figure~\ref{fig:typeIIpred}.
\begin{Schunk}
\begin{Sinput}
R> gpfrPredType2 <- gpfrPredict(train = gpfrFit, testInputGP = x_new, 
+    testTime = t_new, uReg = scalar_new, gpReg = NULL)
R> plot(gpfrPredType2, type = 'prediction')
R> lines(t_new, response_new, type='b', col = 4, pch = 19, cex = 0.6, 
+    lty = 3, lwd = 2)
\end{Sinput}
\end{Schunk}
\vspace{-1cm}


\section{Extensions} \label{sec:concl}

Among the future functionalities of \pkg{GPFDA}, is the Bayesian optimization 
for the covariance function hyper-parameters when the gradient is difficult to obtain. We may also extend the package to deal with non-Gaussian data \citep{wang2014generalized}.

For the NSGPR model, different specifications for the basis functions will be included 
in addition to B-splines. For the MGPR model, the package will be able to model 
the cross-covariance structure between outputs defined on different domains, e.g., 
a time series variable and a spatiotemporal variable.

In addition to the two-step approach that sequentially estimates the mean function (via FR) and covariance functions (via GPR), \pkg{GPFDA} may include an iterative algorithm suggested by \cite{shi2011gaussian} to learn the GPFR model.

\section*{Computational details}

\pkg{GPFDA} uses two auxiliary \proglang{R} packages for visualization: \pkg{fields} \citep{fields} and \pkg{interp} \citep{interp}. For dealing with functional data objects, functionalities of \pkg{fda} \citep{fda} and  \pkg{fda.usc} \citep{fda.usc} are used. \pkg{mvtnorm} \citep{mvtnorm2019} is used for simulating data from a GP in examples shown in \textit{vignettes}. Evaluation of B-spline and cyclic B-spline basis functions is done by using the \proglang{R} packages \pkg{splines} and \pkg{mgcv} \citep{wood2020mgcv}.

The results in this paper were obtained using \proglang{R}~3.6.3. The development repository of \pkg{GPFDA} is hosted on GitHub at \url{https://github.com/gpfda/GPFDA-dev}. \proglang{R} itself and all packages used are available from the Comprehensive \proglang{R} Archive Network (CRAN) at \url{https://CRAN.R-project.org/}.


\bibliography{refs}

\newpage
\begin{appendix}

\section{Gradient and second derivatives} \label{app:derivatives}

The marginal log-likelihood of $\ve{\theta}$ in \eqref{eq:3.lik1} depends on 
$\mathbf{\Psi}(\ve t, \ve t') = \mathbf{K}(\ve t, \ve t') + \sigma_\epsilon^2 \mathbf{I}$, where the $(i,j)$th element of $\mathbf{K}$ is given by
$[\mathbf{K}(\ve t, \ve t')]_{ij} = \COV (f_i, f_j) = k(\ve t_i, \ve t_j')$.

\subsection{Derivatives of the log-likelihood with respect to hyper-parameters} \label{app.cov0}

The gradient of the log-likelihood function is given by
\[
\frac{\partial{l}}{\partial \theta_j} = 
\frac{1}{2}\tr \Big(
(\ve \alpha \ve \alpha^\top - \mathbf{\Psi}^{-1}) \frac{\partial\mathbf{\Psi}}{\partial \theta_j} 
\Big),
\]
where $\ve \alpha = \mathbf{\Psi}^{-1} \ve x$ and $\tr(\ve A)$ denoting the trace of matrix $\ve A$. The second derivatives are
\[
\frac{\partial^2{l}}{\partial \theta_i \theta_j} = 
\frac{1}{2}\tr \Bigg[
(\ve \alpha \ve \alpha^\top - \mathbf{\Psi}^{-1}) 
\Bigg(
\frac{\partial^2\mathbf{\Psi}}{\partial \theta_i \partial \theta_j} - \ve A_{ij}
\Bigg) 
- \ve \alpha \ve \alpha^\top  \ve A_{ij}
\Bigg],
\]
where 
\[
\ve A_{ij} = \frac{\partial\mathbf{\Psi}}{\partial \theta_i} 
\mathbf{\Psi}^{-1}
\frac{\partial\mathbf{\Psi}}{\partial \theta_j}
\]

In Section~\ref{app.cov}, we use different specifications for the kernel 
$k(\cdot,\cdot)$.

\subsection{Derivatives of covariance functions with respect to hyper-parameters} \label{app.cov}

\subsection*{$\bullet$ Linear}
\[
k_{\text{linear}}(\ve t, \ve t') = \exp(a_0) + \sum^Q_{q=1}\exp(a_q) \ve t_{q} \ve t_{q}'.
\]
For $a_0$:
\[
\frac{\partial\mathbf{\Psi}}{\partial a_0}(\ve t, \ve t') = 
\frac{\partial^2\mathbf{\Psi}}{\partial a_0^2}(\ve t, \ve t') = \exp(a_0)
\]
For $a_q, \ q=1,\dots,Q$:
\[
\frac{\partial\mathbf{\Psi}}{\partial a_q}(\ve t, \ve t') = 
\frac{\partial^2\mathbf{\Psi}}{\partial a_q^2}(\ve t, \ve t') = 
\exp(a_q) \ve t_q \ve t_q'
\]

\subsection*{$\bullet$ Powered exponential}
\[
k_{\text{pow.ex}}(\ve t, \ve t') = \exp(v)  \exp \Big(- \sum^Q_{q=1} \exp(w_q)(\ve t_{q}-\ve t_{q}')^\gamma  \Big), \qquad  0 < \gamma \leq 2.
\]
For $v$:
\[
\frac{\partial\mathbf{\Psi}}{\partial v}(\ve t, \ve t') = 
\frac{\partial^2\mathbf{\Psi}}{\partial v^2}(\ve t, \ve t') = 
k_{\text{pow.ex}}(\ve t, \ve t')
\]
For $w_q, \ q=1,\dots,Q$:
\[
\frac{\partial\mathbf{\Psi}}{\partial w_q}(\ve t, \ve t') = 
- k_{\text{pow.ex}}(\ve t, \ve t') \exp(w_q)(\ve t_{q}-\ve t_{q}')^\gamma 
\]
\[
\frac{\partial^2\mathbf{\Psi}}{\partial w_q^2}(\ve t, \ve t') = 
k_{\text{pow.ex}}(\ve t, \ve t') 
\Big(
\exp(2 w_q)(\ve t_{q}-\ve t_{q}')^{2\gamma} - 
\exp(w_q)(\ve t_{q}-\ve t_{q}')^\gamma 
\Big)
\]

\subsection*{$\bullet$ \Matern{} ($\nu = 3/2$)}
We have $d_{(2)} = \sum^Q_{q=1} w_q (\ve t_{q}-\ve t_{q}')^2, \qquad \omega_q \geq 0$ and
\[
k_{\text{matern3/2}}(\ve t, \ve t') = \exp(v) \Big(1 + \sqrt{3} d_{(2)}^{1/2}\Big) \exp \Big(-\sqrt{3}  d_{(2)}^{1/2}\Big).
\]
For $v$:
\[
\frac{\partial\mathbf{\Psi}}{\partial v}(\ve t, \ve t') = 
\frac{\partial^2\mathbf{\Psi}}{\partial v^2}(\ve t, \ve t') = 
k_{\text{matern3/2}}(\ve t, \ve t')
\]
For $w_q, \ q=1,\dots,Q$:
\[
\frac{\partial\mathbf{\Psi}}{\partial w_q}(\ve t, \ve t') = -\frac{3}{2} \exp(v) \exp(w_q) (\ve t_{q}-\ve t_{q}')^2  \exp \Big(-\sqrt{3}  d_{(2)}^{1/2}\Big)
\]
\[
\frac{\partial^2\mathbf{\Psi}}{\partial w_q^2}(\ve t, \ve t') = -\frac{3}{2} \exp(v) \exp(w_q) (\ve t_{q}-\ve t_{q}')^2  \exp \Big(-\sqrt{3}  d_{(2)}^{1/2}\Big)
\Big[
1 - \frac{\sqrt{3}}{2} d_{(2)}^{-1/2} \exp(w_q) (\ve t_{q}-\ve t_{q}')^2
\Big]
\]

\subsection*{$\bullet$ \Matern{} ($\nu = 5/2$)}
\[
k_{\text{matern5/2}}(\ve t, \ve t') = \exp(v) \Big(1 + \sqrt{5} d_{(2)}^{1/2} + \frac{5}{3} d_{(2)}\Big) \exp \Big(-\sqrt{5}  d_{(2)}^{1/2}\Big).
\]
For $v$:
\[
\frac{\partial\mathbf{\Psi}}{\partial v}(\ve t, \ve t') = 
\frac{\partial^2\mathbf{\Psi}}{\partial v^2}(\ve t, \ve t') = 
k_{\text{matern5/2}}(\ve t, \ve t')
\]
For $w_q, \ q=1,\dots,Q$:
\[
\frac{\partial\mathbf{\Psi}}{\partial w_q}(\ve t, \ve t') =  - \frac{5}{6} \exp(v) \exp(w_q) (\ve t_{q}-\ve t_{q}')^2  \exp \Big(-\sqrt{5}  d_{(2)}^{1/2}\Big) 
\big[
1 + \sqrt{5} d_{(2)}^{1/2}
\big]
\]
\[
\frac{\partial^2\mathbf{\Psi}}{\partial w_q^2}(\ve t, \ve t') =  - \frac{5}{6} \exp(v) \exp(w_q) (\ve t_{q}-\ve t_{q}')^2  \exp \Big(-\sqrt{5}  d_{(2)}^{1/2}\Big) 
\big[
1 + \sqrt{5} d_{(2)}^{1/2} - \frac{5}{2} \exp(w_q) (\ve t_{q}-\ve t_{q}')^2
\big]
\]

\subsection*{$\bullet$ Rational quadratic}
\[
k_{\text{rat.qu}}(\ve t, \ve t') = \exp(v) 
\Big(
 1 + \sum^Q_{q=1} \exp(w_q) (\ve t_{q}-\ve t_{q}')^2 
 \Big)^{-\exp(\alpha)} , 
 \ \alpha \geq 0	
\]
For $v$:
\[
\frac{\partial\mathbf{\Psi}}{\partial v}(\ve t, \ve t') = 
\frac{\partial^2\mathbf{\Psi}}{\partial v^2}(\ve t, \ve t') = 
k_{\text{rat.qu}}(\ve t, \ve t')
\]
For $w_q, \ q=1,\dots,Q$:
\[
\frac{\partial\mathbf{\Psi}}{\partial w_q}(\ve t, \ve t') = 
- \exp(\alpha) \exp(v) \Big(
1 + \sum^Q_{q=1} \exp(w_q) (\ve t_{q}-\ve t_{q}')^2 
\Big)^{-\exp(\alpha)-1} \exp(w_q) (\ve t_q - \ve t_q')^2
\]
\begin{align*}
\frac{\partial^2\mathbf{\Psi}}{\partial w_q^2}(\ve t, \ve t')  =  
- \exp(\alpha) \exp(v) 
\Big[ &
(-\exp(\alpha)-1) \Big(
1 + \sum^Q_{q=1} \exp(w_q) (\ve t_{q}-\ve t_{q}')^2 
\Big)^{-\exp(\alpha)-2}\exp(2 w_q) (\ve t_q - \ve t_q')^4  \\
& - \Big(
1 + \sum^Q_{q=1} \exp(w_q) (\ve t_{q}-\ve t_{q}')^2 
\Big)^{-\exp(\alpha)-2}\exp(w_q) (\ve t_q - \ve t_q')^2 
\Big]
\end{align*}
For $\alpha$:
\[
\frac{\partial\mathbf{\Psi}}{\partial \alpha}(\ve t, \ve t') = 
- \exp(\alpha) k_{\text{rat.qu}}(\ve t, \ve t') \log 
\Big(
1 + \sum^Q_{q=1} \exp(w_q) (\ve t_{q}-\ve t_{q}')^2 
\Big)
\]
\[
\frac{\partial^2\mathbf{\Psi}}{\partial \alpha^2}(\ve t, \ve t') = 
- \exp(\alpha)  \Big[ \frac{\partial\mathbf{\Psi}}{\partial \alpha}(\ve t, \ve t') +   k_{\text{rat.qu}}(\ve t, \ve t') \Big] \log 
\Big(
1 + \sum^Q_{q=1} \exp(w_q) (\ve t_{q}-\ve t_{q}')^2 
\Big)
\]

\subsection*{$\bullet$  Noise term}
\[
k_{\text{noise}}(\ve t, \ve t') =  \exp(\sigma_\epsilon^2) I_{\left\{\ve t = \ve t'\right\}}
\]
\[
\frac{\partial\mathbf{\Psi}}{\partial\sigma_\epsilon^2}(\ve t, \ve t')  = 
\frac{\partial^2\mathbf{\Psi}}{\partial(\sigma_\epsilon^2)^2}(\ve t, \ve t')  = 
\exp(\sigma_\epsilon^2)
\]
\end{appendix}


\end{document}